\documentclass[10pt,twocolumn,letterpaper]{article}

\usepackage[]{cvpr}
\usepackage{tikz}
\usepackage[super]{nth}
\usepackage{siunitx}
\usepackage{graphicx}
\usepackage{amsmath}
\usepackage{multirow}
\usepackage{mathtools}
\usepackage{todonotes}
\usepackage{threeparttable}
\usepackage{amssymb}
\usepackage{csquotes}
\usepackage{booktabs}
\usepackage[acronym]{glossaries}
\usepackage{algorithm2e}
\usepackage[capitalize]{cleveref}

\crefname{section}{Sec.}{Secs.}
\Crefname{section}{Section}{Sections}
\Crefname{table}{Table}{Tables}
\crefname{table}{Tab.}{Tabs.}

\def\cvprPaperID{10}
\def\confName{OAGM}
\def\confYear{2023}

\newacronym{pdf}{p.d.f}{probability density function}
\newacronym{ebm}{EBM}{energy based model}
\newacronym{mcmc}{MCMC}{Markov chain Monte Carlo}
\newacronym{sde}{SDE}{stochastic differential equations}
\newacronym{mri}{MRI}{magnetic resonance imaging}
\newacronym{tv}{TV}{total variation}
\newacronym{vn}{VN}{variational network}
\newacronym{ipalm}{iPALM}{inertial proximal alternating linearized minimization}
\newacronym{acl}{ACL}{auto-calibration lines}
\newacronym{psnr}{PSNR}{peak signal-to-noise ratio}
\newacronym{cs}{CS}{compressed sensing}
\newacronym{pdfs}{PDFS}{proton density fat-suppressed}
\newacronym{pd}{PD}{proton density}
\newacronym{vesde}{VE-SDE}{variance exploding stochastic differential equation}
\newacronym{zf}{ZF}{zero-filled}
\newacronym{rss}{RSS}{root sum of squares}

\newcommand{\R}{\mathbb{R}}
\newcommand{\norm}[1]{\lVert#1\rVert}
\DeclareMathOperator*{\argmin}{arg\,min}
\DeclareMathOperator{\prox}{prox}

\newcommand{\crop}{\tikz[thick, scale=0.25]{%
		\draw [shorten <= -0.4](0.25, 0.25) -- ++(-0.7, 0.);%
		\fill[white] (-0.15, 0.15) rectangle (-0.35, 0.35);%
		\draw (-0.25, -0.25) -- ++(0., 0.7);%

		\draw [shorten <= -0.4] (-0.25, -0.25) -- ++(0.7, 0.);%
		\fill[overlay, white] (0.15, -0.15) rectangle (0.35, -0.35);%
		\draw [shorten >= -2](0.25, 0.25) -- ++(0., -0.5);%
}}
\newcommand{\cropsmall}{\tikz[scale=0.15]{%
		\draw [shorten <= -0.4](0.25, 0.25) -- ++(-0.7, 0.);%
		\fill[white] (-0.15, 0.15) rectangle (-0.35, 0.35);%
		\draw (-0.25, -0.25) -- ++(0., 0.7);%

		\draw [shorten <= -0.4] (-0.25, -0.25) -- ++(0.7, 0.);%
		\fill[overlay, white] (0.15, -0.15) rectangle (0.35, -0.35);%
		\draw [shorten >= -2](0.25, 0.25) -- ++(0., -0.3);%
}}

\RestyleAlgo{ruled}
\SetKwInput{KwRequire}{Require}
\SetKwInput{KwDefine}{Define}

\title{Joint Non-Linear MRI Inversion with Diffusion Priors}

\author{%
	Moritz Erlacher, Martin Zach\\
	Graz Univeristy of Technology\\
	Institute of Computer Graphics and Vision\\
	Inffeldgasse 16/II, 8010 Graz\\
	{\tt\small \{moritz.erlacher@student., zach@\}tugraz.at}
}

\begin{document}
\maketitle
\begin{abstract}
	\Gls{mri} is a potent diagnostic tool, but suffers from long examination times.
	To accelerate the process, modern \gls{mri} machines typically utilize multiple coils that acquire sub-sampled data in parallel.
	Data-driven reconstruction approaches, in particular diffusion models, recently achieved remarkable success in reconstructing these data, but typically rely on estimating the coil sensitivities in an off-line step.
	This suffers from potential movement and misalignment artifacts and limits the application to Cartesian sampling trajectories.
	To obviate the need for off-line sensitivity estimation, we propose to jointly estimate the sensitivity maps with the image.
	In particular, we utilize a diffusion model --- trained on magnitude images only --- to generate high-fidelity images while imposing spatial smoothness of the sensitivity maps in the reverse diffusion.
	The proposed approach demonstrates consistent qualitative and quantitative performance across different sub-sampling patterns.
	In addition, experiments indicate a good fit of the estimated coil sensitivities.
\end{abstract}
\glsresetall
\section{Introduction}
\label{sec:intro}
\Gls{mri} provides detailed images of the human anatomy with excellent soft-tissue contrast non-invasively.
However, patient throughput is limited by long examination times, which can be reduced by acquiring less data.
In recent years, reconstruction methods for sub-sampled \gls{mri} have seen a lot of progress.
Classical variational approaches impose prior knowledge --- such as gradient- or wavelet-sparsity~\cite{lustig2007sparse,knoll2011tgv} ---  onto the reconstruction.
In general, such hand-crafted priors fail to accurately model the underlying data distribution~\cite{kobler2022tdv} and purely data-driven approaches now represent state-of-the-art in \gls{mri} reconstruction~\cite{wang2016accelerating,song2022solving,chung2022scorebased,zach2023stable,gungor2023adaptive,hammernik2019ml_mri,jalal2021robust,luo2023bayesian,sriram2020endtoend}.
Methods following a discriminative approach directly map k-space to image-space.
This necessitates data-image pairs, which are scarcely available~\cite{sriram2020endtoend,hammernik2019ml_mri,wang2016accelerating}.
Moreover, such methods do not generalize well to different acquisition modalities without retraining.
In contrast, generative approaches learn the underlying data distribution, relying only on much more abundantly available DICOM data.
In addition, they are able to generalize to different acquisition modalities by adapting the forward model~\cite{zach2023stable,chung2022scorebased,song2022solving,jalal2021robust,luo2023bayesian,gungor2023adaptive}.

As a particular instantiation of generative models, diffusion models have recently gained a lot of interest~\cite{song2019sbm_gradients,chung2022scorebased,chung2022come}.
They convince with high sample quality without adversarial training~\cite{song2020advances}.
On a high level, diffusion models generate samples by gradually transforming a \enquote{simple} distribution into the complex data distribution.
This is typically modelled by \gls{sde}, where sampling from the prior distribution amounts to reversing the \gls{sde} by using the gradient of the log perturbed data distribution learned by a deep neural network.
This gradient is also known as the score function, hence such models are also commonly known as score-based generative models.

\begin{figure*}
	\centering
	\includegraphics[width=0.7\linewidth]{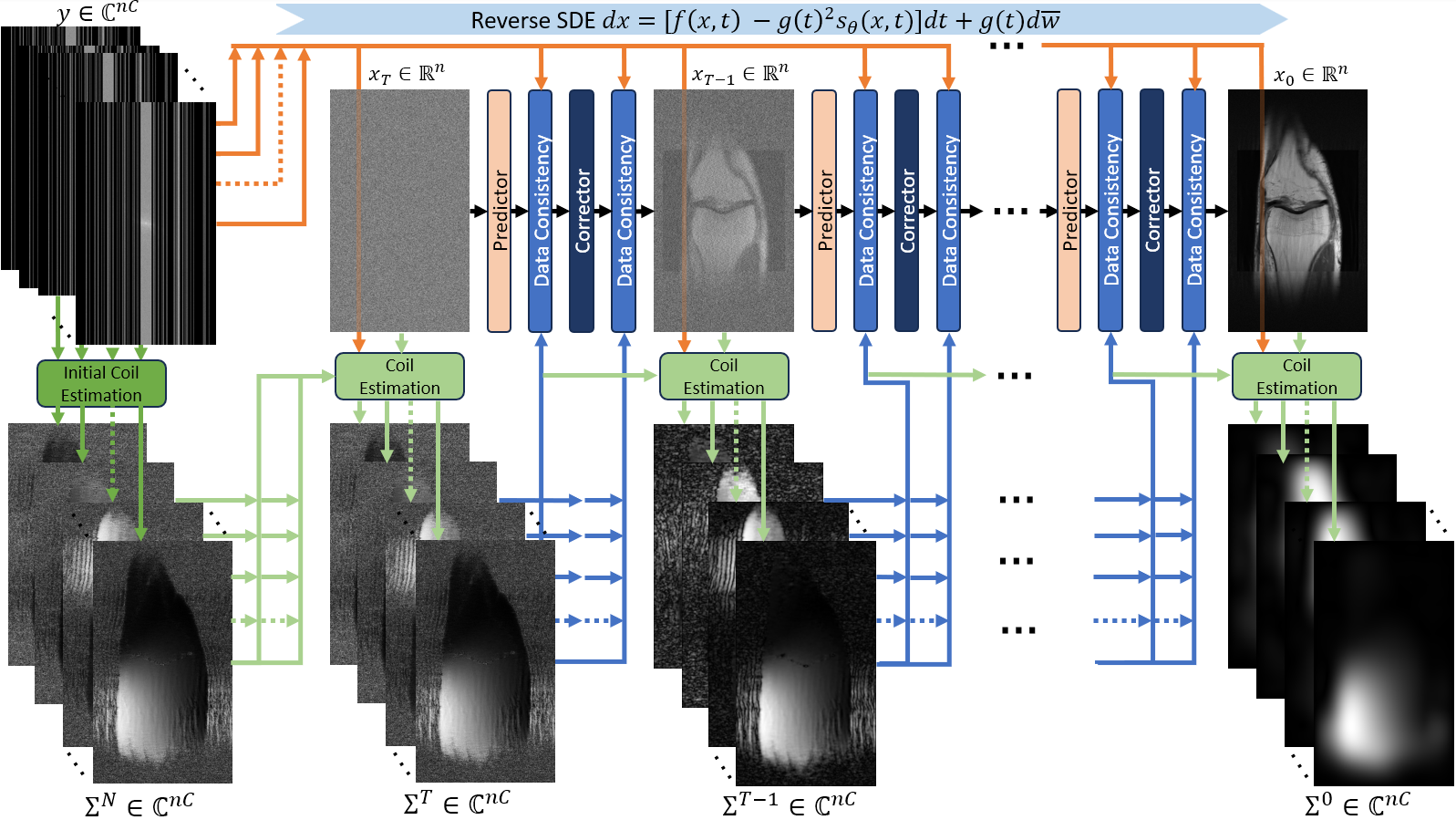}
	\caption{%
		Proposed joint \gls{mri} image reconstruction and coil sensitivity estimation approach.
		In the reverse diffusion, the image and the coil sensitivities are jointly estimated.
	}%
	\label{fig:approach}
\end{figure*}

In this work, we propose to use diffusion models as an implicit prior during joint reconstruction of \gls{mri} images and coil sensitivities.
Our approach is trained on broadly available DICOM data~\cite{zbontar2019fastmri}, resulting in a model that can be used for parallel imaging and different sub-sampling patterns without retraining.
A sketch of our proposed approach is shown in~\cref{fig:approach}.
\subsection{Related work}%
\label{sec:related_work}
Diffusion models for \gls{mri} were proposed by different authors in recent years~\cite{chung2022scorebased,song2022solving,jalal2021robust,luo2023bayesian,gungor2023adaptive}.
To combine the implicit diffusion-prior with the data-likelihood, the authors of~\cite{jalal2021robust} use annealed Langevin dynamics~\cite{song2019sbm_gradients}.
Notably, their work required complex-valued \gls{mri} images for training and relies on off-line sensitivity estimation, e.g.\ using ESPIRiT~\cite{uecker2014espirit}.

Off-line sensitivity estimation is prone to motion and misalignment artifacts, and not trivially applicable for non-Cartesian sampling trajectories~\cite{zach2023stable,knoll2011tgv,ying2007jsense}.
To avoid off-line sensitivity estimation,~\cite{chung2022scorebased} propose to apply a single score function --- trained on reference \gls{rss} reconstructions --- to the real and imaginary parts of the individual coil images.
Thus, the number of gradient evaluations needed in their algorithm is proportional to the number of acquisition coils.
The authors also propose an alternative that relies on off-line sensitivity estimation, which suffers from the same shortcomings mentioned before.
Joint image reconstruction and coil sensitivity estimation were first proposed by~\cite{ying2007jsense}, who explicitly parametrized the sensitivities with low-order polynomials and used alternating minimization to solve the resulting optimization problem.
~\cite{zach2023stable} instead enforce spatial smoothness on the coil sensitivities during the optimization with \gls{ipalm}~\cite{Pock2016ipalm}, and utilize a \gls{ebm} resembling the data distribution to get high-fidelity reconstructions.
However, \gls{ebm} training is known to be unstable and requires hand tuning of many parameters~\cite{nijkamp2020anatomy,zach2023stable}.
 
In this work, we propose a joint reconstruction algorithm that leverages an implicit prior given by a diffusion model.
In contrast to~\cite{chung2022scorebased}, our algorithm requires only one gradient evaluation of the diffusion model in one iteration of the reverse diffusion.
In addition, we propose a novel way to utilize a diffusion model for reconstruction problems of arbitrary image size, where only a cropped region follows the data distribution learned by the diffusion model.
\section{Background}%
\label{sec:background}
This paper is built on two main pillars: Diffusion models and joint non-linear \gls{mri} inversion.
In this section, we will briefly introduce these concepts, but refer the reader to the provided references for more details.
\subsection{Diffusion models}
Diffusion models circumvent the computation of the (typically intractable) partition function arising in maximum-likelihood density estimation by instead estimating the gradient of the log-prior, $\nabla \log p_{X_0}$, which is referred to as the \emph{score}.
To facilitate efficient sampling and to accurately model low-density regions, the authors of~\cite{song2020sbm} propose to construct an \gls{sde}
\begin{equation}
	\mathrm{d}X = f(X, t)\,\mathrm{d}t + g(t)\,\mathrm{d}w
	\label{eq:sde}
\end{equation}
where \( w \) is the standard Wiener process, \( f : \R^n \times [0, \infty) \to \R^n \) is the drift and \( g : [0, \infty) \to \R \) is the diffusion coefficient.
In this work we choose \( f \equiv 0 \) and define \( g(t) = \sqrt{\dot{\sigma}^2(t)} \) (the choice of \( \sigma : [0, T] \to \R \) is detailed in~\cref{sec:implementation_details}), which is known as the \emph{variance exploding} \gls{sde} and has close connections to classical heat diffusion~\cite{zach2023explicit}.
Denoting with \( X_t \) the random variable obeying~\eqref{eq:sde}, the \emph{score matching} objective reads
\begin{equation}
	\min_\theta \tilde{\mathbb{E}} \bigl[\gamma(t)\norm{\nabla_1 \log p_{X_t\mid X_0}(x_t, x_0) - s_{\theta}(x_t, t)}^2_2 / 2 \bigr].
	\label{eq:background:loss_sde_sbm}
\end{equation}
Here, \( \tilde{\mathbb{E}}[\,\cdot\,] \) denotes \( \int_0^T \mathbb{E}_{x_0\sim p_{X_0},x_t \sim p_{X_t\mid X_0}(\,\cdot\,,x_0)} [\,\cdot\,]\ \mathrm{d}t \), \( s_\theta : \R^n \times [0, T] \to \R^n \) is the diffusion model and $\gamma : [0, T] \to \R_+$ is a weighting function.
\( T > 0 \) is an artificial time horizon that can be set to \( T = 1 \) without loss of generality.
In our setup, \( \nabla_1 \log p_{X_t \mid X_0}(x_t, x_0) = \frac{x_t - x_0}{\sigma^2(t)} \) and we choose \( \gamma(t) = \sigma^2(t) \).
For more detail about the training procedure, we refer to~\cite{song2020sbm}.

To generate samples from the data distribution, we run the reverse time \gls{sde}
\begin{equation}
	\mathrm{d}X = [f(X, t) - g^2(t)\nabla \log p_{X_t}(X)]\,\mathrm{d}t + g(t)\,\mathrm{d} \bar{w}
\end{equation}
starting from \( x_{T} \sim p_{X_T} \) until \( t = 0 \), where we use the learnt score model \( s_\theta(\,\cdot\,, t) \) in place of \( \nabla \log p_{X_t} \).
With our choice of \( f \) and \( g \), a straight forward time discretization of this process yields
\begin{equation}
	x_i \leftarrow x_{i+1} + (\sigma^2_{i+1} - \sigma^2_i)s_{\theta}(x_{i+1}, \sigma_{i+1}) + \sqrt{\sigma^2_{i+1} - \sigma^2_i}z
\end{equation}
where $z \sim \mathcal{N}(0,I)$.
\subsection{Non-linear \gls{mri} inversion}%
\label{sec:background:solving_mri_with_sbms}
In this work we assume the acquisition model
\begin{equation}
	y = \mathcal{A}(x, \Sigma) + \epsilon
	\label{eq:background:forward_model}
\end{equation}
where the data $y \in \mathbb{C}^{nC}$ are acquired through the non-linear measurement operator 
\begin{equation}
	\begin{aligned}
		\mathcal{A} : \R^n \times \mathbb{C}^{nC} &\to \mathbb{C}^{nC}\\
		(x, \Sigma) &\mapsto \begin{pmatrix}
			\mathcal{F}_{\Omega}(c_1 \odot x \oslash |\Sigma|_{\mathcal{C}})\\
			\mathcal{F}_{\Omega}(c_2 \odot x \oslash |\Sigma|_{\mathcal{C}})\\
			\vdots\\
			\mathcal{F}_{\Omega}(c_C \odot x \oslash |\Sigma|_{\mathcal{C}})
		\end{pmatrix}
	\end{aligned}
	\label{eq:background:forward_operator_parallel}
\end{equation}
acting on the underlying image $x \in \R^n$ with \( \epsilon \in \mathbb{C}^{nC} \) summarizing the additive acquisition noise.
In the above, the shorthand \( \Sigma \coloneqq (c_j)^C_{j=1} \in \mathbb{C}^{nC} \) denotes the sensitivity maps of the \( C \in \mathbb{N} \) coils and \( |\,\cdot\,|_{\mathcal{C}} : \mathbb{C}^{nC} \to \R^n_+ \) denotes the \gls{rss} map \( (c_j)^C_{j=1} \mapsto \sqrt{\sum_{j=1}^C |c_j|^2} \) where \( |\,\cdot\,| \) is the complex modulus acting element-wise on its argument (see~\cite{zach2023stable} on why the division with $|\Sigma|_{\mathcal{C}}$ is necessary).
Further, $\mathcal{F}_{\Omega} : \mathbb{C}^n \to \mathbb{C}^{\bar{n}}$ is the (possibly non-uniform) Fourier transform acquiring the spectrum at locations specified by the trajectory $\Omega$.
For the sake of simplicity, we only consider the case where \( \mathcal{F}_\Omega = MF \), where \( F : \mathbb{C}^n \to \mathbb{C}^n \) is the standard Fourier transform on the Cartesian grid and \( M \) is a binary diagonal matrix specifying the acquired frequencies (hence also \( n = \bar{n} \)).

Motivated by recent advances in non-linear inversion, in this work we tackle the reconstruction by jointly estimating the image with the coil sensitivities.
In detail, let \( D : \R^n \times \mathbb{C}^{nC} \to \R_+ \) denote the least-squares objective of~\eqref{eq:background:forward_model}, i.e. 
\begin{equation}
	D : (x, \Sigma) \mapsto \norm{\mathcal{A}(x, \Sigma) - y}^2_2/2
	\label{eq:background:least_squares}
\end{equation}
The optimization problem \( \argmin_{(x, \Sigma)} D(x, \Sigma) \) is highly underspecified due to ambiguities between \( x \) and \( \Sigma \) in~\eqref{eq:background:forward_operator_parallel}.
In addition, reconstructed images would exhibit strong sub-sampling artifacts.
We resolve the ambiguities by imposing a hand-crafted smoothness prior on the coil sensitivities and utilize the implicit prior provided by a diffusion model to generate high fidelity reconstructions.
We discuss the details in the next section.
\section{Methods}
For the reconstruction of the \gls{mri} image we follow the predictor-corrector sampling introduced by~\cite{song2020sbm,chung2022scorebased}.
To ensure data consistency during the reverse diffusion, similar to~\cite{chung2022scorebased}, we utilize gradient updates of the form
\begin{equation}
	x_{i} \leftarrow x_{i+1} - \lambda_{i+1} \nabla_1 D(x_{i+1}, \Sigma_{i+1}).
	\label{eq:background:data_fidelity}
\end{equation}
In detail, let \( \mathcal{A}|_\Sigma : \R^n \to \mathbb{C}^{nC} : x \mapsto \mathcal{A}(x, \Sigma) \) denote the linearization of \( \mathcal{A} \) in the first argument around \( \Sigma \).
Then,
\begin{equation}
	\nabla_1 D(x, \Sigma) = (\mathcal{A}|_{\Sigma})^\ast(\mathcal{A}(x, \Sigma) - y)
\end{equation}
with
\begin{equation}
	\begin{aligned}
		(\mathcal{A}|_{\Sigma})^\ast : \mathbb{C}^{nC} &\to \R^n\\
		(y_j)_{j=1}^C &\mapsto \operatorname{Re}\biggl(\sum_{j=1}^C \mathcal{F}_{\Omega}^{-1} (y_j) \odot \bar{c}_j \oslash |\Sigma|_C\biggr).
	\end{aligned}
\end{equation}
denoting the adjoint of \( \mathcal{A}|_{\Sigma} \) and $\lambda_i \in [0,1]$ is the step size (see~\cite{chung2022scorebased} on why it is restricted to \( [0, 1] \)).

To apply the diffusion model trained on \( \tilde{n} = 320 \times 320 \) images to the data of resolution $n = 640 \times w, w \in \{368,372\}$, we propose the following.
For the input of the diffusion model, we center-crop the image to $\tilde{n}=320 \times 320$ (denoted by a $\crop$ in the superscript in~\cref{alg:background:sbd_sampling}).
After the reverse diffusion update steps, we have found it beneficial to pad the result with the image
\begin{equation}
	x^{\text{fsde}}_i = |F_\Omega^{-1}(y)|_\mathcal{C} + \sigma^2_iz
\end{equation}
satisfying the forward \gls{sde} instead of the result of the gradient step on the data fidelity.
The operator $\operatorname{pad} : \R^{\tilde{n}} \times \R^n \to \R^n$ in~\cref{alg:background:sbd_sampling} implements this padding.
\subsection{Estimating coil sensitivities during reverse diffusion}
In addition to regularizing the image, we also estimate the coil sensitivities during the reverse diffusion process.
In particular, we assume that the sensitivity maps are smother than the imaged anatomy.
To enforce smoothness, we closely follow~\cite{zach2023stable}.
In detail, during the iterations of their proposed algorithm, they smooth the individual coil sensitivities by
\begin{equation}
	\prox_{\mu \tilde{B}} : c_j \mapsto (Q_{\mu} \circ \operatorname{Re})(c_j) + \imath (Q_{\mu} \circ \operatorname{Im})(c_j)
\end{equation}
where \( Q_\mu : x \mapsto \mathcal{S}^{-1} \bigl( \operatorname{diag}(\xi_i + \mu)^{-1} \mathcal{S}(\mu x) \bigr) \) utilizes the discrete sine transform \( \mathcal{S} \) and \( \xi_i = 2 - 2\cos \phi_i \) are the eigenvalues of the discrete Laplace operator for equally spaced angles \( \phi_i = \frac{\pi i}{n} \) for \( i = 0,\dotsc,n-1 \) (see~\cite[Chap. 19.4]{numerical92} for more detail).
In the above, \( \mu > 0 \) defines the strength of smoothing and \( \imath \) is the imaginary unit.
Notice that this can be interpreted as the proximal operator of a quadratic gradient penalization
\begin{equation}
	\tilde{B} : c_j \mapsto \frac{1}{2} \bigl(\norm{\mathrm{D}\operatorname{Re}(c_j)}^2_2 + \norm{\mathrm{D}\operatorname{Im}(c_j)}^2_2\bigr)
\end{equation}
where \( \mathrm{D} : \R^n \to \R^{2n} \) is the discrete gradient operator (see e.g.~\cite{chambolle2010first}).
Let \( B : (c_j)^C_{j=1} \mapsto \sum_{j=1}^C \tilde{B}(c_j) \), then by proximal calculus rules
\begin{equation}
	\prox_{\mu B}(\Sigma) = (\prox_{\mu \tilde{B}}(c_1), \dotsc, \prox_{\mu \tilde{B}}(c_C))^\top.
\end{equation}
The update step for the coil sensitivities can thus be summarized as
\begin{equation}
  \Sigma_i \leftarrow \operatorname{prox}_{\mu_{i + 1} B}(\Sigma_{i+1} - \mu_{i + 1} \nabla_2 D(x_{i + 1}, \Sigma_{i+1}))
\end{equation}
where the gradient step on \( D \)
\begin{equation}
	(\nabla_2 D(x, \Sigma))_j = \biggl( \frac{\kappa_j}{|\Sigma|_\mathcal{C}} - \frac{\alpha_j}{|\Sigma|^3_\mathcal{C}} \biggr) \odot x
\end{equation}
ensures data consistency and the proximal step enforces smoothness.
In the above, \( \kappa_j = \mathcal{F}_\Omega^{-1} (s_j) \), \( \alpha_j = c_j \odot \bigl( \sum_{E \in \{ \operatorname{Re}, \operatorname{Im} \}} E(c_j) \odot E(s_j) \bigr) \) with \( s_j = \mathcal{F}_\Omega(x \odot c_j \oslash |\Sigma|_\mathcal{C}) - y_j \) denoting the residual of the \( j\)-th channel.
We initialize the coil sensitivities with the \gls{zf} estimate
\begin{equation}
	c_j = \frac{\mathcal{F}_{\Omega}^{-1}(y_j)}{|\mathcal{F}_\Omega^{-1}(y)|_\mathcal{C}}.
\end{equation}
The algorithm is summarized in~\cref{alg:background:sbd_sampling}.
 
\begin{algorithm}[t]
	\DontPrintSemicolon
	\KwRequire{$s_{\theta}, M, N, \{\sigma_i\}, \{\lambda_i\}, \{x_{i}^{\text{fsde}}\}, \{\mu_i\}$}
	\KwResult{\( x_0 \), \( \Sigma_0 \)}
	$\Sigma_N = \frac{\mathcal{F}_{\Omega}^{-1}(y)}{|\mathcal{F}_\Omega^{-1}(y)|_\mathcal{C}}$\\
	$\Sigma_N = \frac{\Sigma_N}{\norm{\Sigma_N}^2_2}$\\
	\For{\( i \leftarrow N - 1,\dotsc,0\)}{
		$z \sim \mathcal{N}(0, I)$\\
		$x_i \leftarrow \operatorname{pad}(x_{i+1}^{\crop} + (\sigma^2_{i+1} - \sigma^2_i)s_{\theta}(x_{i+1}^{\crop}, \sigma_{i+1}) + \sqrt{\sigma^2_{i+1} - \sigma^2_i}z, x_{i + 1}^{\text{fsde}})$\\
		$x_i \leftarrow x_i - \lambda_{i + 1} \nabla_1 D(x_{i + 1}, \Sigma_{i + 1})$\\
		\For{\(j \leftarrow 1,\dotsc,M\)}{
			$z \sim \mathcal{N}(0,I)$\\
			$\epsilon_i \leftarrow 2r^2\norm{z}_2^2 / \norm{s_{\theta}(x^{j-1,\crop}_{i + 1}, \sigma_{i + 1})}_2^2$\\
			$x^{j,\crop}_i \leftarrow x^{j-1,\crop}_i + \epsilon_is_{\theta}(x^{j-1,\crop}_{i + 1}, \sigma_{i + 1}) + \sqrt{2\epsilon_i}z$\\
		}
		\( x_i \leftarrow \operatorname{pad}(x^{M,\crop}_{i + 1}, x_{i + 1}^{\text{fsde}}) \)\\
		$x_i \leftarrow x_{i + 1} - \lambda_{i + 1} \nabla_1D(x_{i + 1}, \Sigma_{i + 1})$\\
		$\Sigma_i \leftarrow \operatorname{prox}_{\mu_{i + 1} B}(\Sigma_{i+1} - \mu_{i + 1} \nabla_2 D(x_{i + 1}, \Sigma_{i+1}))$\\
	}
	\caption{Diffusion-based joint \gls{mri} image reconstruction and coil sensitivity estimation.}
	\label{alg:background:sbd_sampling}
\end{algorithm}
\subsection{Implementation details}\label{sec:implementation_details}
The model we use in this work follows the model of~\cite{song2020sbm} and ~\cite{chung2022scorebased}, both following a U-Net model architecture~\cite{ronneberger2015unet}.
We use four BigGAN~\cite{brock2019biggan} residual blocks with additional skip connections for the latent vector and a self-attention block at the smallest scale.
For each block a bias is added conditioned with Gaussian Fourier projections of the current time step $t \in [0, T]$, resulting in an embedding of size $128\times1\times1$.
We employ $\{64, 64, 128, 128\}$ feature maps in the corresponding up and down sampling blocks.
Our network has \num{11951041} trainable parameters, in comparison to~\num{61433601} in the work of~\cite{chung2022scorebased}.

For the training, we closely follow~\cite{chung2022scorebased} and~\cite{song2020sbm}.
We optimizing the objective~\cref{eq:background:loss_sde_sbm} for \num{10000} epochs using a batch size of \num{3} with Adam~\cite{kingma2014adam} ($\beta_1=\num{0.9}$, $\beta_2 = \num{0.999}$, learning rate \num{e-4}).
As proposed by~\cite{song2020advances}, exponential moving average is used during training with a momentum of \num{0.999}.
For the noise variance schedule, we use the geometric series $\sigma(t) = \sigma_{\text{min}}(\frac{\sigma_{\text{max}}}{\sigma_{\text{min}}})^t$, with $\sigma_{\text{min}} = 0.01$ and $\sigma_{\text{max}} = \num{378}$.
Training was done on a NVIDIA TITAN V with 12GB of memory resulting in 20 days of training.\\
For sampling, we set $N = \num{1000}$, $M = \num{1}$. The choice of \( r \) follows~\cite{song2020sbm,chung2022scorebased} with $r=0.0075$.
For $\lambda$ and $\mu$ an exponential schedule of the form $\chi_i=e^{\zeta_i}$ is used where $\zeta_i$ is equispaced between $\log\chi_{N}$ and $\log \chi_{1}$ for $\chi \in \{\lambda, \mu\}$.
We choose an exponentially decreasing schedule for $\lambda$ to prioritize stronger data consistency influence at the beginning of the diffusion process, when noise dominates, and lower its impact towards the end to mitigate the introduction of sub-sampling artifacts.
Similarly an increasing exponential schedule is choosen for $\mu$ to ensure that the coils are not initially influenced by noise during reconstruction.
The parameters were found by grid search and are shown in~\cref{tab:best_parameters}.
\begin{table}
	\resizebox{\columnwidth}{!}{
		\begin{threeparttable}
			\caption{%
				Parameters $\lambda$ and $\mu$ found by grid search.
			}%
			\label{tab:best_parameters}
			\begin{tabular}{lccccccccc}
				\toprule
				   &     \multirow{2}{*}{Acc.} & \multicolumn{2}{c}{$\lambda_{N}$} & \multicolumn{2}{c}{$\lambda_{1}$} & \multicolumn{2}{c}{$\mu_{N}$}         & \multicolumn{2}{c}{$\mu_{1}$}        \\\cmidrule(lr{.75em}){3-4} 	\cmidrule(lr{.75em}){5-6} 	\cmidrule(lr{.75em}){7-8} 	\cmidrule(lr{.75em}){9-10}
				   &                                & PD         & PDFS     & PD         & PDFS     & PD             & PDFS         & PD             & PDFS         \\ \midrule
				Cartesian& $4$                                & 0.56            & 0.40          & 0.07            & 0.05          & $\num{e-6}$       & $\num{e-6}$     & 25       & 25  \\
				Cartesian\tnote{*} & $4$                      & 0.80            & 0.80          & 0.30            & 0.30          & $\num{e-6}$       & $\num{e-6}$     & 25       & 10  \\
				Gaussian &$4$                                 & 0.56            & 0.56          & 0.21            & 0.07          & $\num{e-6}$       & $\num{e-6}$     & 25       & 25  \\
				Radial& $11$                                  & 0.70            & 0.70          & 0.21            & 0.21          & $\num{e-6}$       & $\num{e-6}$     & 20       & 20  \\
				\bottomrule
			\end{tabular}
			\begin{tablenotes}
				\item[*] Swapped phase encoding direction
			\end{tablenotes}
		\end{threeparttable}
	}
\end{table}
\subsection{Experimental data}
The training and test data is taken from the fastMRI dataset~\cite{zbontar2019fastmri}.
For training our model we used the \gls{rss} reconstructions of size $320 \times 320$ available in the dataset from both the coronal \gls{pd} and \gls{pdfs} contrasts.
To avoid training with noise, we focus on the central 10 slices, resulting in $973 \times 10 = \num{9730}$ images.
For consistent intensity ranges during training, the intensities in each image were normalized to lie in $[0, 1]$.
Due to computational limitations, we restrict the test set to $15$ \gls{pd} and $15$ \gls{pdfs} randomly chosen central slices from the fastMRI validation set (see Appendix).
\subsection{Comparison and evaluation}
We compare our approach to the joint non-linear inversion method presented in~\cite{zach2023stable} using a Charbonnier-smoothed \gls{tv}~\cite{charbonnier1997tv} regularizer and the end-to-end \gls{vn} from~\cite{sriram2020endtoend}.
Due to the fact, that the approach of~\cite{chung2022scorebased} does not work with the full size data we could not include it in this comparison.
The \gls{vn} was trained on the CORPD training split of the fastMRI dataset with random 4-fold Cartesian sub-sampling patterns using $\qty{8}{\percent}$ \gls{acl}.
We quantitatively compare the reconstructions using \gls{psnr}:
Since the reconstructed images vary strongly in magnitude, we define the \gls{psnr} as \( 10\log_{10} \frac{\tilde{n}\norm{x^{\text{gt}}}_\infty^2}{\norm{x^{\cropsmall} - x^{\text{gt}}}_2^2} \) where \( x^{\text{gt}} \) is the \( 320 \times 320 \) reference center region and \( x^{\crop} \) is the center region of the reconstruction.
We evaluate the estimated coil sensitivities qualitatively by inspecting the \gls{rss} null-space residual~\cite{uecker2014espirit}.
As in~\cite{zach2023stable}, our approach necessitates a post-processing of the signal intensities, as low-intensity regions (e.g.\ air) are systematically under-estimated.
To remedy this, we follow~\cite{zach2023stable} and fit a spline curve on the scatter of reference and reconstruction intensities (separately for \gls{pd} and \gls{pdfs} contrast) on an independent hold-out set containing \num{3} images (see Appendix).
\section{Results}
Figure~\ref{fig:results} shows qualitative results of reconstruction for different sub-sampling patterns. In the first row, \num{4}-fold Cartesian sub-sampling using \qty{8}{\percent} \gls{acl} is shown.
The \gls{tv} reconstructions already appear over-smoothed, even though the artifacts are still partly present.
The \gls{vn} performs the best in this case, which is expected since it was specifically trained for this task in a discriminative fashion.
For both contrasts, our approach can reproduce high-frequency details, although there are slight sub-sampling artifacts visible.
This might indicate sub-optimal parameter choices.

The second row shows the same sub-sampling pattern but with swapped phase encoding direction.
\gls{tv} removes the artifacts at the cost of high-frequency details.
\gls{vn} does not perform well on this task, which indicates that this model does not adapt to different sub-sampling patterns.
Our approach removes the artifacts and reproduces the high-frequency details.

The third row shows \num{4}-fold Gaussian sub-sampling. \gls{tv} reproduces very similar \gls{psnr} results, the fogginess is not removed.
Similar to before, the \gls{vn} can not satisfactorily reconstruct the image with other sub-sampling patterns than the one it was trained on.
Our approach removes all sub-sampling artifacts and reproduces a high amount of detail.

The last row shows radial sub-sampling with \num{45} spokes, corresponding to an approximate acceleration factor of \num{11}. 
The \gls{tv} reconstruction does not show any artifacts, but due to the high acceleration factor most details are lost.
The \gls{vn} reproduces some details, but also hallucinates artificial structures and does not fully remove the sub-sampling artifacts.
Similar to the Gaussian sub-sampling, our approach completely removes the smoothness and artifacts and reproduces details well.
\begin{figure*}
  \centering
	\includegraphics[width=0.9\textwidth]{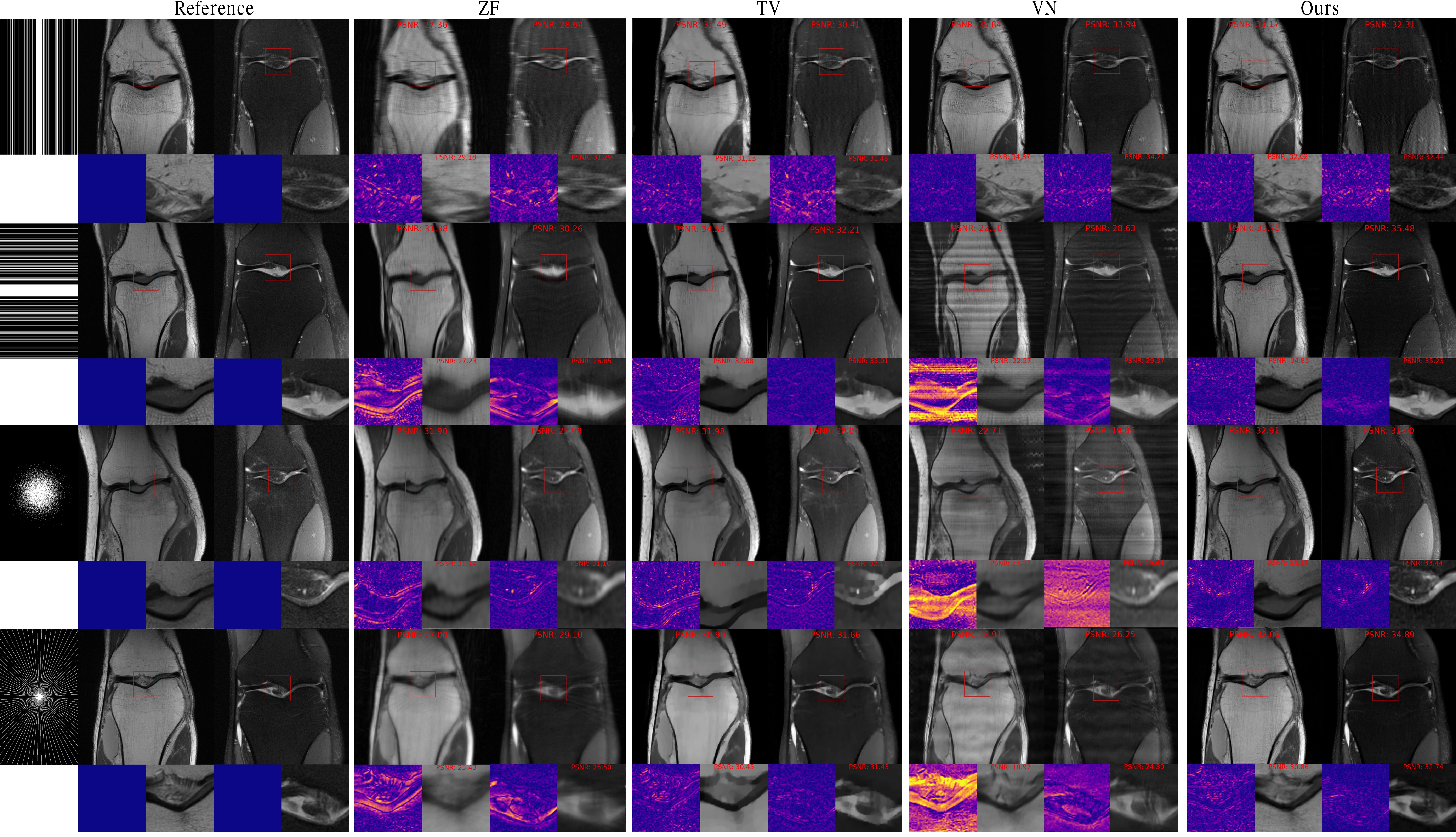}
	\caption{%
		Reconstruction results for the different methods:
		\nth{1} row: \num{4}-fold Cartesian sub-sampling using \qty{8}{\percent} \gls{acl}.
		\nth{2} row: Swapped phase encoding direction.
		\nth{3} row: \num{4}-fold Gaussian sub-sampling.
		\nth{4} row: Radial sub-sampling with \num{45} spokes (\( \approx \num{11} \) acceleration factor).
		The inlays show a zoom of the reconstruction (right) and the magnitude of its difference to the reference (left).
	}%
	\label{fig:results}
\end{figure*}
These results are also reflected in the quantitative analysis shown in~\cref{tab:results}.
With the exception of the Cartesian sub-sampling (on which the \gls{vn} was trained), our method performs best.
\begin{table}
	\resizebox{\columnwidth}{!}{
		\begin{threeparttable}
			\caption{%
				Quantitative results for different sub-sampling patterns using \acrfull{zf}, \acrfull{tv}, \acrfull{vn} and our approach.
				Bold typeface indicates the best method.%
			}%
			\label{tab:results}
			\begin{tabular}{lccccccccc}
				\toprule
				   &     \multirow{2}{*}{Acc.} & \multicolumn{2}{c}{\gls{zf}} & \multicolumn{2}{c}{\gls{tv}} & \multicolumn{2}{c}{\gls{vn}}         & \multicolumn{2}{c}{Ours}        \\\cmidrule(lr{.75em}){3-4} 	\cmidrule(lr{.75em}){5-6} 	\cmidrule(lr{.75em}){7-8} 	\cmidrule(lr{.75em}){9-10}
				   &                                & \gls{pd}         & \gls{pdfs}     & \gls{pd}         & \gls{pdfs}     & \gls{pd}             & \gls{pdfs}         & \gls{pd}             & \gls{pdfs}         \\ \midrule
				Cartesian& $4$                                & 26.67            & 27.97          & 31.35            & 31.18          & \textbf{36.85}       & \textbf{33.72}     & 32.11                & 32.04           \\
				Cartesian\tnote{*} & $4$                     & 30.37             & 28.65          & 32.43            & 30.88          & 23.89                & 29.68              & \textbf{34.34}       & \textbf{32.88}  \\
				Gaussian &$4$                                 & 30.24            & 27.68          & 30.91            & 30.90          & 22.36                & 27.35              & \textbf{31.01}       & \textbf{31.55}  \\
				Radial& $11$                                  & 27.93            & 26.70          & 30.66            & 30.70          & 19.97                & 25.55              & \textbf{32.18}       & \textbf{31.78}  \\
				\bottomrule
			\end{tabular}
			\begin{tablenotes}
				\item[*] Swapped phase encoding direction
			\end{tablenotes}
		\end{threeparttable}
	}
\end{table}

The proposed method takes approximately \qty{140}{\second} per image.
On the other hand, \gls{tv} and \gls{vn} complete the reconstruction in approximately \qty{3}{\second} and \qty{0.2}{\second} respectively.
The reconstruction time is largely determined by the number of steps \( N \) taken in the reverse diffusion.
Reducing the number of iterations to \( N = \num{200} \) cuts the reconstruction time down to approximately \qty{30}{\second} and only has a limited impact on the result.
We believe that other acceleration techniques, such as the one presented in~\cite{chung2022come} are readily applicable to our approach.

\begin{figure}
  \centering
	\includegraphics[width=0.9\linewidth]{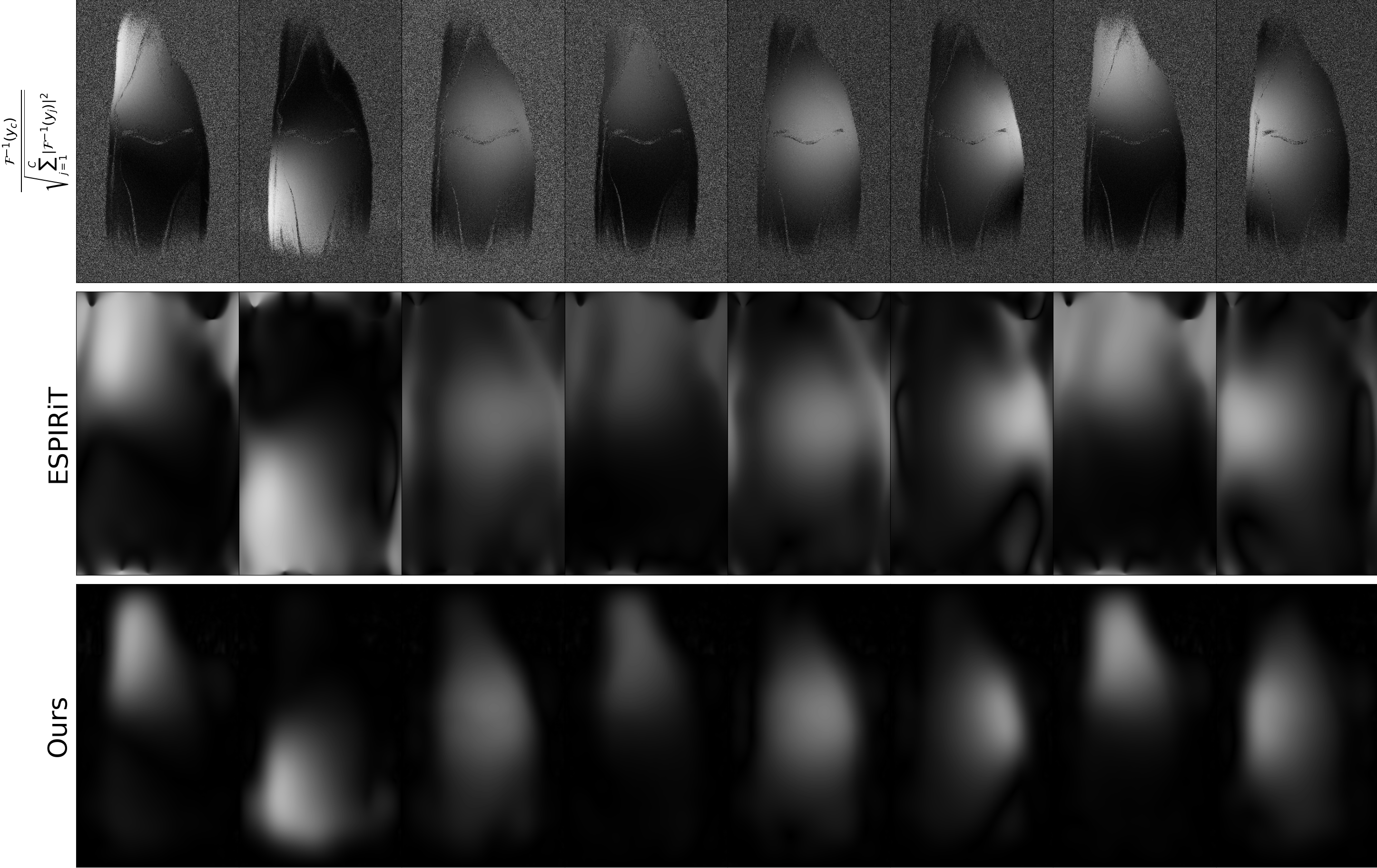}
	\caption{%
		Reference coil sensitivities computed from the fully-sampled data (top), ESPIRiT~\cite{uecker2014espirit} estimation (middle), and the result of our joint estimation (bottom).
		These are the results for the image in the first row of~\cref{fig:results} (\gls{pd} contrast).
	}%
	\label{fig:coils_estimate}
\end{figure}

\begin{figure}
	\centering
	\begin{subfigure}{0.49\linewidth}
		\centering
		\includegraphics[width=1.0\linewidth]{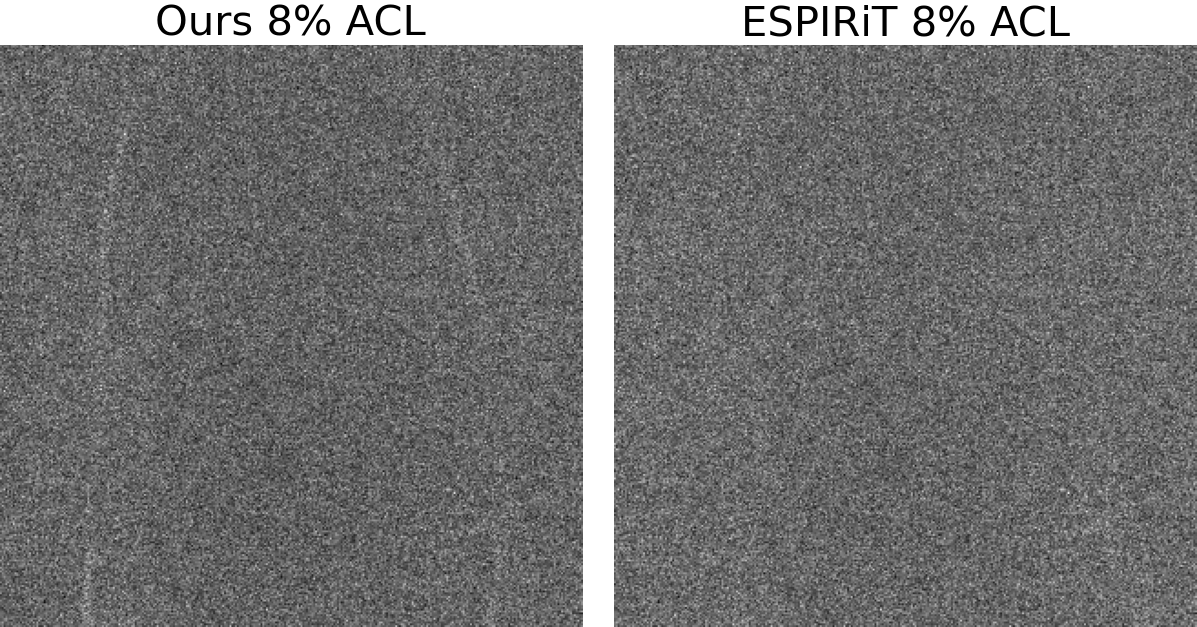}%
		\label{fig:null_space_8ACL}
	\end{subfigure}
	\begin{subfigure}{0.49\linewidth}
		\centering
		\includegraphics[width=1.0\linewidth]{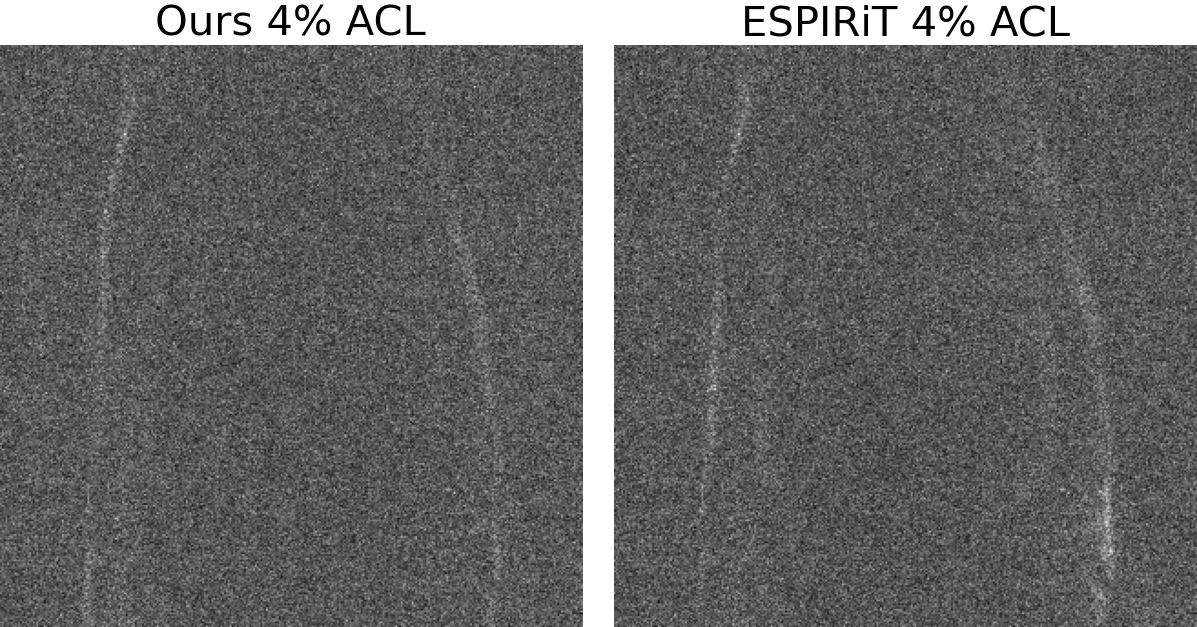}%
		\label{fig:null_space_4ACL}
	\end{subfigure}
	\caption{%
		\gls{rss} null-space residual of our coil sensitivities versus ESPIRiT~\cite{uecker2014espirit}.
		These are the results for the image in the first row of~\cref{fig:results} (\gls{pd} contrast) using $8\%$ and $4\%$ \gls{acl}.
	}
	\label{fig:null_space}
\end{figure}

To evaluate the estimated coil sensitivities, we compare our approach with ESPIRiT~\cite{uecker2014espirit}.
The results are shown in~\cref{fig:coils_estimate} using the first image in~\cref{fig:results} for $\times 4$ Cartesian sub-sampling.
It can be seen that our approach is able to estimate the coil sensitivities adequately, by comparing them to the reference.
In addition to qualitatively assessing the coil sensitivities, we evaluate the goodness-of-fit in terms of the forward model by visualizing the \gls{rss} null-space residual~\cite{uecker2014espirit}.
Any residual signal components indicate a sub-optimal fit.
In~\cref{fig:null_space}, we compare the coil sensitivity estimation on \num{4}-fold Cartesian sub-sampling using \qty{8}{\percent} and \qty{4}{\percent} \gls{acl} between ESPIRiT and our approach.
We see that ESPIRiT achieves slightly better results when subjected to \qty{8}{\percent} \gls{acl}, whereas our approach performs better for \qty{4}{\percent} \gls{acl}.
This is in line with our expectation, since the joint estimation can utilize all available data, whereas ESPIRiT is restricted to the \gls{acl}.
\section{Conclusion}
We propose a joint non-linear inversion algorithm for \gls{mri} that leverages the implicit prior defined by a diffusion model.
Our approach is capable of reconstructing \gls{mri} images satisfactorily for different sub-sampling patterns, maintaining qualitative and quantitative performance.
In contrast to the method presented in~\cite{chung2022scorebased}, our approach only requires one evaluation of the score function in each step of the reverse diffusion, irrespective of the number of coils, leading to faster reconstructions.
Additionally, we propose a novel way of applying diffusion models to data of different size, where only a sub-region follows the data distribution implicit in the diffusion model.
In addition to reconstructing the image, our approach also estimates the coil sensitivities.
Experiments show that the estimation is as good as, or sometimes superior to, classical off-line estimation methods, but does not suffer from their drawbacks.

Expanding upon this research, one potential direction is finding a method using the full size data without padding using the forward \gls{sde}. Further reducing the 
reconstruction time is crucial, by applying different sampling approach or incorporating the ideas of ~\cite{chung2022come}, the speed gap could be reduced. 
{\small
	\bibliographystyle{ieee_fullname}
	\bibliography{egbib}
}
\appendix
\section{Experimental data}
\begin{table}
	\caption{Test data files}%
	\label{tab:test_files}
	\begin{tabular}{cc}
		PDFS & PD\\
		\toprule
		\texttt{file1000593.h5} & \texttt{file1002187.h5} \\
		\texttt{file1002451.h5} & \texttt{file1001893.h5} \\
		\texttt{file1001440.h5} & \texttt{file1001298.h5} \\
		\texttt{file1000591.h5} & \texttt{file1000976.h5} \\
		\texttt{file1001851.h5} & \texttt{file1000625.h5} \\
		\texttt{file1001338.h5} & \texttt{file1002546.h5} \\
		\texttt{file1002159.h5} & \texttt{file1001221.h5} \\
		\texttt{file1000990.h5} & \texttt{file1000477.h5} \\
		\texttt{file1000538.h5} & \texttt{file1001585.h5} \\
		\texttt{file1002214.h5} & \texttt{file1001077.h5} \\
		\texttt{file1000858.h5} & \texttt{file1000635.h5} \\
		\texttt{file1001834.h5} & \texttt{file1001184.h5} \\
		\texttt{file1000273.h5} & \texttt{file1001163.h5} \\
		\texttt{file1000818.h5} & \texttt{file1001064.h5} \\
		\texttt{file1002067.h5} & \texttt{file1000660.h5} \\
	\end{tabular}
\end{table}

\begin{table}
	\caption{Validation data files}%
	\label{tab:validation_files}
	\begin{tabular}{cc}
		PDFS & PD\\
		\toprule
		\texttt{file1001365.h5} & \texttt{file1000759.h5} \\
		\texttt{file1001191.h5} & \texttt{file1001759.h5} \\
		\texttt{file1000280.h5} & \texttt{file1002570.h5} \\
	\end{tabular}
\end{table}
For the experiments, we used the following files from the fastMRI dataset~\cite{zbontar2019fastmri}.
The test data files in~\cref{tab:test_files} were used for the quantitative and qualitative results in the paper.
The validation files in~\cref{tab:validation_files} were used for the spline fit.
\begin{figure}
	\centering
	\begin{subfigure}{0.49\textwidth}
		\centering
		\includegraphics[width=1.0\linewidth]{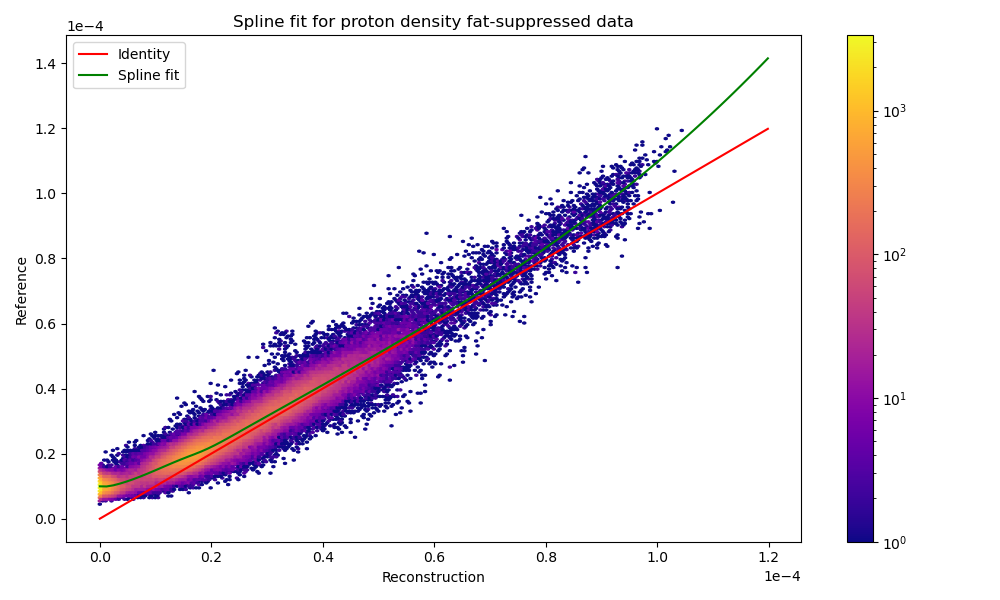}%
		\label{fig:spline_fs}
	\end{subfigure}
	\begin{subfigure}{0.49\textwidth}
		\centering
		\includegraphics[width=1.0\linewidth]{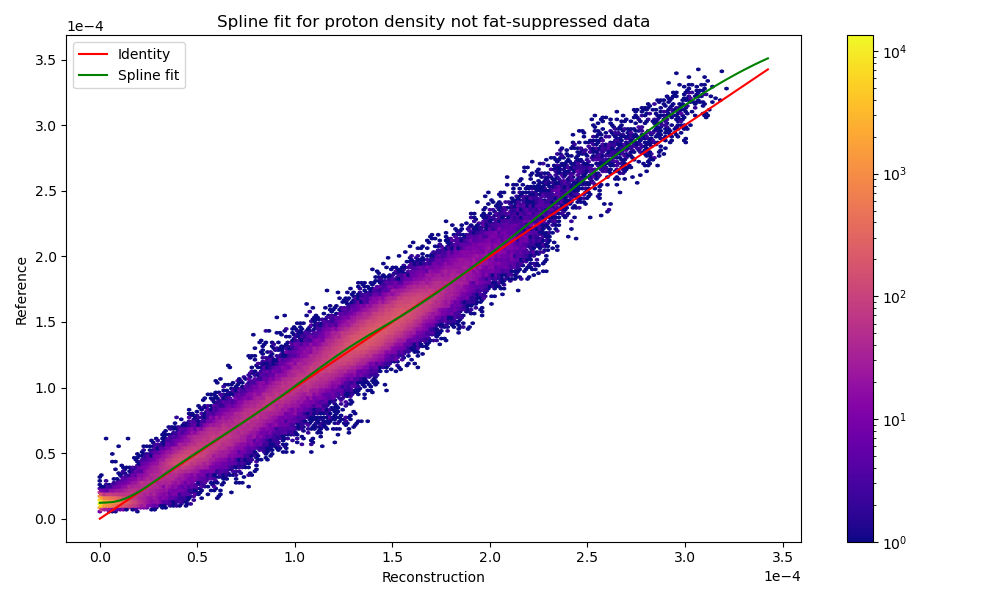}%
		\label{fig:spline_non_fs}
	\end{subfigure}
	\caption{%
		Spline fit on the \gls{pdfs} data (top). Spline fit on the \gls{pd} data (bottom).
	}%
	\label{fig:spline}
\end{figure}
\section{Spline fit}
\cref{fig:spline} shows the spline fit performed on the validation data of the coronal \gls{pdfs} and \gls{pd} contrast.
\end{document}